\documentclass[12pt]{article}
\usepackage{amssymb}

\makeatletter
\def\numberbysection{\@addtoreset{equation}{section}
 	\def\theequation{\thesection.\arabic{equation}}}
\makeatother

\numberbysection

\newcommand{\be}{\begin{eqnarray}}
\newcommand{\ee}{\end{eqnarray}}
\newcommand{\non}{\nonumber}
\newcommand{\tr}{\mathop{\rm tr}\nolimits}
\newcommand{\id}{\mathbb{I}}
\newcommand{\csch}{\mathop{\rm csch}\nolimits}
\newcommand{\sech}{\mathop{\rm sech}\nolimits}

\newcommand{\ch}{\mathop{\rm ch}\nolimits}
\newcommand{\sh}{\mathop{\rm sh}\nolimits}
\newcommand{\tnh}{\mathop{\rm th}\nolimits}
\newcommand{\cth}{\mathop{\rm cth}\nolimits}

\begin{document}

\begin{titlepage}
\strut\hfill
\vspace{.5in}
\begin{center}

\LARGE The solution of an open XXZ chain\\
\LARGE with arbitrary spin revisited\\[1.0in]
\large Rajan Murgan\footnote{email:rmurgan@svsu.edu} and Chris Silverthorn\footnote{e-mail: cbsilver@svsu.edu}\\[0.8in]
\large Department of Physics,\\ 
\large Saginaw Valley State University,\\ 
\large 7400 Bay Road University Center,  MI 48710 USA\\

\end{center}

\vspace{.5in}

\begin{abstract}
The Bethe ansatz solutions for an open XXZ spin chain with arbitrary spin with $N$ sites and nondiagonal boundary terms are revisited. The anisotropy parameter, for cases
considered here, has values $\eta = i \pi {r\over q}$, where $r$ and $q$ are positive integers with $q$ restricted to odd integers. Numerical results are presented to support the solutions.  
\end{abstract}
\end{titlepage}

\setcounter{footnote}{0}

\section{Introduction}\label{sec:intro}

Over the years, significant progress has been made on the solutions of the integrable open spin-$1/2$ XXZ quantum spin chain, in particular those with general or nondiagonal 
boundary terms. These works resulted in a number of approaches and methods which have been utilized in the solutions of the spectrum of the model. They include solutions obtained via 
certain functional relations obeyed by the transfer matrix of the model \cite{BR} for various special cases, such as that with boundary parameters obeying certain 
contraints \cite{CLSW}-\cite{YZ1} or that with root of unity values for the anisotropy parameter \cite{MNS3}. Works on the open XXZ spin chain such as that 
based on the deformed Onsager algebra \cite{BK, BK2, BK3} and solutions from functional relations based on Yang-Baxter algebra for the open XXZ model with 
general boundary terms \cite{Galleas} have also shed light on the nature of the solutions obtained for the open XXZ quantum spin chain. Further, an interesting method based
on (generalized) coordinate Bethe ansatz has been used in solving the open XXZ spin chain (among others) with nondiagonal boundaries in \cite{CRS, CRS2}.
More recent works based on the separation of variables method \cite{Niccoli, FKN, KMG} and the $\it{inhomogeneous}$ T-Q equation approach \cite{CYSW} have served 
as important advancements to understand this crucial model.     

Extension of the spin-$1/2$ solutions of the XXZ spin chain to include arbitrary spin $s$ $\Large (s = \frac{1}{2}, 1, \frac{3}{2},\ldots\Large)$ have also received 
considerable interest. In \cite{Doikou}, the Bethe ansatz solution for the open XXZ spin chain with alternating spins was constructed utilizing the method of 
\cite{Nep}. This solution relies on a certain constraint among the boundary parameters. In \cite{FNR}, a generalization of this constraint was found as well 
as a second set of Bethe ansatz equations necessary to obtain all the eigenvalues. One application of such a solution of the open XXZ chain with arbitrary spin
is, the $s=1$ case enables one to investigate the boundary version of the supersymmetric sine-Gordon model \cite{BSSG}-\cite{BSSG3}. In particular, the Bethe ansatz 
solutions of the open spin-$1$ XXZ chain have been used to derive the nonlinear integral equations for the supersymmetric sine-Gordon model \cite{ANS, MurganSSG}.

Motivated by these studies, we revisit solutions of an open spin-$s$ XXZ spin chain studied in \cite{Murganspins}. We extend the anisotropy parameter values to 
include $\eta = i \pi {r\over q}$, where $r$ and $q$ are positive integers with $q$ assuming odd integer values. To avoid duplication, only the irreducible fractions for 
$\frac{r}{q}$ are considered. Part of the motivation to consider this problem is that to the extent that a number can be 
approximated by a rational number, this should in principle extend the solutions in \cite{Murganspins} to include a larger class of imaginary values of $\eta$ than 
was presented earlier. In addition, to our knowledge, Bethe ansatz solution for open spin-$s$ XXZ spin chain for the case considered here has not been given before. As in
\cite{Murganspins}, we consider cases with at most two arbitrary boundary parameters.
In the crucial work on open spin-$s$ given in \cite{FNR}, Bethe ansatz solution presented there works when certain constraint are obeyed by the boundary parameters. Also, although 
in a recent important advance \cite{CYSW}, the solution for arbitrary values of $\eta$ and the boundary parameters is given (with an
unconventional term in the Bethe equations) for the open XXZ chain, such a solution has been written down only for $s=\frac{1}{2}$ case. 

This paper is arranged as follows: In Sec. 2, the transfer matrix of the open XXZ spin chain \cite{Sklyanin} is briefly reviewed. In addition, functional relations
that the transfer matrices obey for $\eta = i\pi\frac{r}{q}$ are reviewed. This is followed by the derivation and presentation of the Bethe ansatz solutions for the open XXZ spin chain model with 
arbitrary spin for cases with two arbitrary boundary parameters in Sec. 3. These solutions are restricted to roots of unity values of the anisotropy parameter, 
$\eta = i \pi {r\over q}$, where $r$ and $q$ are positive integers (with $q$ assuming any odd positive integers). Numerical results 
are given in Sec. 4, using $s = \frac{1}{2}$ and $s = 1$ as examples to support and to check for the completeness of the solutions given (presence of all $(2s+1)^N$ eigenvalues). We do this for selected 
values of number of sites $N$, $\eta$ and the boundary parameters. Some concluding remarks and potential future works follow in Sec. 5.

\section{Transfer matrices and functional relations}\label{sec:transfer}

We present here a brief review on the commuting transfer matrices for $N$-site open XXZ quantum spin chain. The
spin-$1/2$ transfer matrix $t^{(\frac{1}{2},\frac{1}{2})}(u)$, whose auxiliary space as well as each of its $N$ quantum spaces are two-dimensional
is given by \cite{Sklyanin}
\be
t^{(\frac{1}{2},\frac{1}{2})}(u) = \tr_{0} K^{+}_{0}(u)\  
T_{0}(u)\  K^{-}_{0}(u)\ \hat T_{0}(u)\,,
\label{transfer}
\ee
where the trace is taken over the ``auxiliary space'' 0. $T_{0}(u)$ and $\hat T_{0}(u)$ are the monodromy matrices given by
\be
T_{0}(u) = R_{0N}(u) \cdots  R_{01}(u) \,,  \qquad 
\hat T_{0}(u) = R_{01}(u) \cdots  R_{0N}(u) \,,
\label{monodromy}
\ee

The $R$ matrix is given by
\be
R(u) = \left( \begin{array}{cccc}
	\sinh  (u + \eta) &0            &0           &0            \\
	0                 &\sinh  u     &\sinh \eta  &0            \\
	0                 &\sinh \eta   &\sinh  u    &0            \\
	0                 &0            &0           &\sinh  (u + \eta)
\end{array} \right) \,,
\label{bulkRmatrix}
\ee 
where $\eta$ is the bulk anisotropy parameter; and $K^{\mp}(u)$ are
$2 \times 2$ matrices whose components
are given by \cite{dVGR, GZ}
\be
K_{11}^{-}(u) &=& 2 \left( \sinh \alpha_{-} \cosh \beta_{-} \cosh u +
\cosh \alpha_{-} \sinh \beta_{-} \sinh u \right) \non \\
K_{22}^{-}(u) &=& 2 \left( \sinh \alpha_{-} \cosh \beta_{-} \cosh u -
\cosh \alpha_{-} \sinh \beta_{-} \sinh u \right) \non \\
K_{12}^{-}(u) &=& e^{\theta_{-}} \sinh  2u \,, \qquad 
K_{21}^{-}(u) = e^{-\theta_{-}} \sinh  2u \,,
\label{Kminuscomponents}
\ee
and
\be
K^{+}(u) = \left. K^{-}(-u-\eta)\right\vert_{(\alpha_-,\beta_-,\theta_-)\rightarrow
(-\alpha_+,-\beta_+,\theta_+)} \,,
\ee 
where $\alpha_{\mp} \,, \beta_{\mp} \,, \theta_{\mp}$ are the boundary
parameters.

Using the fusion procedure \cite{fusion}, one can similarly construct the open chain transfer matrix $t^{(j,s)}(u)$,
whose auxiliary space is ($2j+1$)-dimensional and each of its $N$ quantum spaces are ($2s+1$)-dimensional \cite{fusion2, fusion3, Zhou}.
The transfer matrix has the commutativity property for $j , j' \in \{\frac{1}{2}, 1,
\frac{3}{2}, \ldots \}$ and any $s \in \{\frac{1}{2}, 1, \frac{3}{2},
\ldots \}$,
\be
\left[ t^{(j,s)}(u) \,, t^{(j',s)}(u') \right] = 0 \,.
\label{commutativity}
\ee
Furthermore, they also obey the fusion hierarchy \cite{fusion2, fusion3},
\be
t^{(j-\frac{1}{2},s)}(u- j\eta)\, t^{(\frac{1}{2},s)}(u) =
t^{(j,s)}(u-(j-\frac{1}{2})\eta)  + \delta^{(s)}(u-\eta)\,
t^{(j-1,s)}(u-(j+\frac{1}{2})\eta) \,, 
\label{hierarchy}
\ee
In (\ref{hierarchy}), $j = 1,\frac{3}{2},\ldots$. In addition, $t^{(0,s)}=1$, and
$\delta^{(s)}(u)$ is given by
\be
\delta^{(s)}(u) &=& \left[\prod_{k=0}^{2s-1}\xi(u+(s-k+\frac{1}{2})\eta)\right]^{2N} 
{\sh(2u) \sh(2u+4\eta)\over \sh(2u+\eta) \sh(2u+3\eta)}\non \\
&\times& 2^{4}\sh(u+\alpha_{-}+\eta)\sh(u-\alpha_{-}+\eta)\ch(u+\beta_{-}+\eta)\ch(u-\beta_{-}+\eta)\non \\
&\times& \sh(u+\alpha_{+}+\eta)\sh(u-\alpha_{+}+\eta)\ch(u+\beta_{+}+\eta)\ch(u-\beta_{+}+\eta) \,.
\label{dd}
\ee
where $\xi(u) = \sh(u+\eta)\sh(u-\eta)$. 
To avoid any unnecessary repetition, we urge the readers to refer to \cite{FNR} where the details on such a construction can be found. 

Next, we review the $q-$th order functional relations \cite{Nep, Nep2}, the ``fundamental'' transfer matrix, $t^{(\frac{1}{2},s)}(u)$ 
(as well as each of the corresponding eigenvalues, $\Lambda^{(\frac{1}{2},s)}(u)$) obeys for bulk anisotropy values $\eta = i\pi\frac{r}{q}$, where $r$ and $q$ are 
positive integers. The functional relations take the following form,
\be
\lefteqn{t^{(\frac{1}{2},s)}(u) t^{(\frac{1}{2},s)}(u +\eta) \ldots t^{(\frac{1}{2},s)}(u + (q-1) \eta)} \non \\
&-& \delta^{(s)} (u-\eta) t^{(\frac{1}{2},s)}(u +\eta) t^{(\frac{1}{2},s)}(u +2\eta) 
\ldots t^{(\frac{1}{2},s)}(u + (q-2)\eta) \non \\
&-& \delta^{(s)} (u) t^{(\frac{1}{2},s)}(u +2\eta) t^{(\frac{1}{2},s)}(u +3\eta)
\ldots t^{(\frac{1}{2},s)}(u + (q-1) \eta) \non \\
&-& \delta^{(s)} (u+\eta) t^{(\frac{1}{2},s)}(u) t^{(\frac{1}{2},s)}(u +3\eta) t^{(\frac{1}{2},s)}(u +4\eta) 
\ldots t^{(\frac{1}{2},s)}(u + (q-1) \eta) \non \\
&-& \delta^{(s)} (u+2\eta) t^{(\frac{1}{2},s)}(u) t^{(\frac{1}{2},s)}(u +\eta) t^{(\frac{1}{2},s)}(u +4\eta) 
\ldots t^{(\frac{1}{2},s)}(u + (q-1) \eta) - \ldots \non \\
&-& \delta^{(s)} (u+(q-2)\eta) t^{(\frac{1}{2},s)}(u) t^{(\frac{1}{2},s)}(u +\eta) 
\ldots t^{(\frac{1}{2},s)}(u +  (q-3)\eta) \non \\
&+& \ldots  = f(u) \,.
\label{funcrltn}
\ee 
For example, for $q=3$ and $q=5$, the functional relations are
\be
& & t^{(\frac{1}{2},s)}(u) t^{(\frac{1}{2},s)}(u+\eta) t^{(\frac{1}{2},s)}(u+2\eta)
- \delta^{(s)}(u-\eta) t^{(\frac{1}{2},s)}(u+\eta) 
- \delta^{(s)}(u) t^{(\frac{1}{2},s)}(u+2\eta)\non \\
& &\quad - \delta^{(s)}(u+\eta) t^{(\frac{1}{2},s)}(u)
= f(u) \,.
\label{funcrltn2}
\ee
and
\be 
& & t^{(\frac{1}{2},s)}(u) t^{(\frac{1}{2},s)}(u+\eta) t^{(\frac{1}{2},s)}(u+2\eta)t^{(\frac{1}{2},s)}(u+3\eta) t^{(\frac{1}{2},s)}(u+4\eta)\non \\  
& &\quad + \delta^{(s)}(u+\eta) \delta^{(s)}(u-2\eta) t^{(\frac{1}{2},s)}(u) 
+ \delta^{(s)}(u) \delta^{(s)}(u+2\eta) t^{(\frac{1}{2},s)}(u+4\eta)\non \\ 
& &\quad + \delta^{(s)}(u+\eta) \delta^{(s)}(u-\eta) t^{(\frac{1}{2},s)}(u+3\eta) 
- \delta^{(s)}(u+\eta) t^{(\frac{1}{2},s)}(u) t^{(\frac{1}{2},s)}(u+3\eta) t^{(\frac{1}{2},s)}(u+4\eta)\non \\  
& &\quad + \delta^{(s)}(u) \delta^{(s)}(u-2\eta) t^{(\frac{1}{2},s)}(u+2\eta)
- \delta^{(s)}(u) t^{(\frac{1}{2},s)}(u+2\eta) t^{(\frac{1}{2},s)}(u+3\eta) t^{(\frac{1}{2},s)}(u+4\eta)\non \\
& &\quad + \delta^{(s)}(u-\eta) \delta^{(s)}(u+2\eta) t^{(\frac{1}{2},s)}(u+\eta)
- \delta^{(s)}(u+2\eta) t^{(\frac{1}{2},s)}(u) t^{(\frac{1}{2},s)}(u+\eta) t^{(\frac{1}{2},s)}(u+4\eta)\non \\
& &\quad - \delta^{(s)}(u-2\eta) t^{(\frac{1}{2},s)}(u) t^{(\frac{1}{2},s)}(u+\eta) t^{(\frac{1}{2},s)}(u+2\eta)\non \\
& &\quad - \delta^{(s)}(u-\eta) t^{(\frac{1}{2},s)}(u+\eta) t^{(\frac{1}{2},s)}(u+2\eta) t^{(\frac{1}{2},s)}(u+3\eta)
= f(u) \,.
\label{funcrltn4}
\ee
respectively.
 
We note here that the scalar function $f(u)$ that appears in the functional relations above, particularly for even $r$ values, differ slightly from that given in 
\cite{Murganspins} where only the $r = 1$ case was considered. This minor difference in form however, is crucial in order for the functional
relations to be obeyed by the ``fundamental'' transfer matrix. We present below the scalar function $f(u)$ separately for odd $r$ and even $r$ cases respectively (when $q$ assumes any
positive odd integer values): 

Case 1. r = positive odd integers

\be
f_{0}(u)  = \left\{ 
\begin{array}{ll}
    (-1)^{N+1} 2^{-4 s (q-1) N}\sh^{4sN} \left( q u \right)\,, \\
    \qquad \qquad s= {1\over 2}\,, {3\over 2}\,, {5\over 2}\,, \ldots \\
   (-1)^{N+1} 2^{-4 s (q-1) N} \ch^{4sN} \left( q u \right)\,, \\
\qquad \qquad s= 1\,, 2\,, 3\,, \ldots \\
\end{array} \right.
\label{f0}
\ee
and
\be
f_{1}(u) &=& (-1)^{N+1} 2^{5-2 q} \Big( \non \\
& & \hspace{-0.2in}
\sh \left( q \alpha_{-} \right)\ch \left( q \beta_{-} \right)
\sh \left( q \alpha_{+} \right)\ch \left( q \beta_{+} \right)
\ch^{2} \left( q u \right) \non \\
&-&
\ch \left( q \alpha_{-} \right)\sh \left( q \beta_{-} \right)
\ch \left( q \alpha_{+} \right)\sh \left( q \beta_{+} \right)
\sh^{2} \left( q u \right) \non \\
&-&
(-1)^{N} \ch \left( q(\theta_{-}-\theta_{+}) \right)
\sh^{2} \left( q u \right) \ch^{2} \left( q u \right) 
\Big) \,,
\label{f1}
\ee
for $s = {1\over 2}\,, {3\over 2}\,, {5\over 2}\ldots$ 
and 
\be
f_{1}(u) &=& (-1)^{N+1} 2^{5-2 q} \Big( \non \\
& & \hspace{-0.2in}
\sh \left( q \alpha_{-} \right)\ch \left( q \beta_{-} \right)
\sh \left( q \alpha_{+} \right)\ch \left( q \beta_{+} \right)
\ch^{2} \left( q u \right) \non \\
&-&
\ch \left( q \alpha_{-} \right)\sh \left( q \beta_{-} \right)
\ch \left( q \alpha_{+} \right)\sh \left( q \beta_{+} \right)
\sh^{2} \left( q u \right) \non \\
&-&
 \ch \left( q(\theta_{-}-\theta_{+}) \right)
\sh^{2} \left( q u \right) \ch^{2} \left( q u \right) 
\Big) \,,
\label{f1ints}
\ee
for $s = 1\,, 2\,, 3\ldots$.

Case 2. r = positive even integers

\be
f_{0}(u)  = (-1)^{N+2} 2^{-4 s (q-1) N}\sh^{4sN} \left( q u \right)\,,
\label{f0reven}
\ee
and
\be
f_{1}(u) &=& (-1)^{N+1} 2^{5-2 q} \Big( \non \\
& & \hspace{-0.2in}
\sh \left( q \alpha_{-} \right)\ch \left( q \beta_{-} \right)
\sh \left( q \alpha_{+} \right)\ch \left( q \beta_{+} \right)
\ch^{2} \left( q u \right) \non \\
&-&
\ch \left( q \alpha_{-} \right)\sh \left( q \beta_{-} \right)
\ch \left( q \alpha_{+} \right)\sh \left( q \beta_{+} \right)
\sh^{2} \left( q u \right) \non \\
&+&
\ch \left( q(\theta_{-}-\theta_{+}) \right)
\sh^{2} \left( q u \right) \ch^{2} \left( q u \right) 
\Big) \,,
\label{f1reven}
\ee
for $s = {1\over 2}\,,1\,,{3\over 2}\,,2\,,{5\over 2}\ldots$.   

\section{Bethe ansatz}

Here we give the main result of this paper. Essentially, we revisit the Bethe ansatz solutions derived earlier in \cite{Murganspins} for spin-$s$ XXZ chain
with nondiagonal boundary terms, using the method in \cite{Nep, Nep2}, 
for cases with two arbitrary boundary parameters, namely any two from the $\{\alpha_{\pm}\,,\beta_{\pm}\}$ set, e.g. $\{\alpha_{+},\beta_{-}\}$, $\{\alpha_{+},\alpha_{-}\}$, etc. 
We also set $\theta_{-} = \theta_{+}= \theta$, 
where $\theta$ is arbitrary. The remaining boundary parameters are set to some fixed values. The reason behind these choices will be given below for all 
the cases treated here. Readers are also urged to refer to Section 3.1 in \cite{Murganspins}, where such a discussion was also presented.  
In \cite{Murganspins}, we considered the case where $\eta = \frac{i\pi}{q}$, where $q$ assumed odd positive integer values. 
We note that solutions presented here, while bear resemblance to that given in \cite{Murganspins}, are worth reporting since they 
include a wider class of (imaginary) values of the anisotropy parameter $\eta (=i\pi\frac{r}{q})$ that we did not consider before, where $\frac{r}{q}$ refers to 
an irreducible fraction. 

\subsection{Case 1 : One arbitrary $\beta$ and one arbitrary $\alpha$}

As the first case, we consider the solution for an open XXZ quantum spin chain with nondiagonal terms, where the arbitrary
boundary parameters consist of one of the $\beta$'s and one of the $\alpha$'s from the $\{\alpha_{\pm}\,,\beta_{\pm}\}$ set.
The remaining ones are fixed, e.g., if $\beta_{-}\,,\alpha_{-}$ are arbitrary, then $\beta_{+} = \eta$, $\alpha_{+} = \frac{i\pi}{2}$
or other similar combinations. We set $\theta_{-} = \theta_{+} = $ arbitrary. The functional relation method used in this paper was proposed by Nepomechie in \cite{Nep, Nep2} 
to solve the spin-$1/2$ case, which in turn was used in \cite{Murganspins} to study the spin-$s$ case for $\eta = \frac{i\pi}{q}$. When the functional relation 
(\ref{funcrltn}) is expressed as the vanishing determinant of a certain matrix $\cal M$ (following \cite{BR}), one finds that (\ref{funcrltn}) can be written as,
\be
\det {\cal M} = 0 \,,
\label{det}
\ee
where ${\cal M}$ is given by the $q \times q$ matrix
\be
{\cal M} = \left(
\begin{array}{cccccccc}
    \Lambda^{(\frac{1}{2},s)}(u) & -h(u) & 0  & \ldots  & 0 & -h(-u+p \eta)  \\
    -h(-u) & \Lambda^{(\frac{1}{2},s)}(u+p\eta) & -h(u+p \eta)  & \ldots  & 0 & 0  \\
    \vdots  & \vdots & \vdots & \ddots 
    & \vdots  & \vdots    \\
   -h(u+p^{2} \eta)  & 0 & 0 & \ldots  & -h(-u-p(p-1) \eta) &
    \Lambda^{(\frac{1}{2},s)}(u+p^{2}\eta)
\end{array} \right)  \,,
\label{calMalpha}
\ee   
where $p+1 = q$. In the matrix above, successive rows are obtained by simultaneously shifting $u \mapsto u + p \eta$
and cyclically permuting the columns to the right provided that there exists a function $h(u)$ with the following properties
\be
h(u + 2 i \pi) = h \left(u +2 q\eta \right) &=& h(u) \,, \label{cond0} \\
h(u+(q+1)\eta)\ h(-u-(q+1)\eta) &=& \delta^{(s)}(u) \,, \label{cond1} \\
\prod_{j=0}^{q-1} h(u+2j\eta) + \prod_{j=0}^{q-1} h(-u-2j\eta) &=& f(u) 
\,. \label{cond2} 
\ee
Equations (\ref{cond0})-(\ref{cond2}) reduce the problem of finding $h(u)$ to solving the following quadratic equation in $z(u)$,
\be
z(u)^{2} -z(u) f(u) +  \prod_{j=0}^{q-1} \delta^{(s)}\left(u+(2j-1)\eta\right) = 0 \,,
\label{quadratic}
\ee
where 
\be
z(u) = \prod_{j=0}^{q-1} h(u+2j\eta) \,.
\label{producth1}
\ee
Our choice of boundary parameters as mentioned at the beginning of this section makes the discriminants of the corresponding
quadratic equations to be perfect squares. Thus the factorizations such as (\ref{producth1}) can be readily carried out. On the contrary, when all boundary 
parameters are arbitrary, the discriminant is no longer a perfect square, and factoring the result becomes a formidable challenge. Solving the quadratic 
equation (\ref{quadratic}) for $z(u)$, after making use of (\ref{dd}) and (\ref{f0})-(\ref{f1reven}), we obtain the following for $h(u)$,
\be
h(u) &=& (-1)^{2sN} 4\left[\prod_{k=0}^{2s-1}\sh(u+(s-k+\frac{1}{2})\eta)\right]^{2N} 
{\sh(2u+2\eta)\over \sh(2u+\eta)}\non\\
&\times&  \ch u\ch(u-\eta) (\sh u+(-1)^{2sN}i\ch\beta)\sh(u-\alpha){\ch\left({1\over 2}(u+\alpha+\eta) \right)\over 
 \ch\left({1\over 2}(u-\alpha-\eta)\right)}
\label{hoddr}
\ee
for odd integer values of $r$ and
\be
h(u) &=& -4\left[\prod_{k=0}^{2s-1}\sh(u+(s-k+\frac{1}{2})\eta)\right]^{2N} 
{\sh(2u+2\eta)\over \sh(2u+\eta)}\non\\
&\times&  \ch u\ch(u-\eta) (\sh u-i\ch\beta)\sh(u-\alpha){\ch\left({1\over 2}(u+\alpha+\eta) \right)\over 
 \ch\left({1\over 2}(u-\alpha-\eta)\right)}
\label{hevenr}
\ee
for even integer values of $r$. 

The structure of the matrix ${\cal M}$ (\ref{calMalpha}), suggests that its 
null eigenvector has the form $\big( Q(u)\,, Q(u+(q-1)\eta) \,, \ldots
\,, Q(u+(q-1)^{2}\eta) \big)$, where $Q(u)$ has the periodicity property
\be
Q(u + 2i\pi) = Q(u) \,.
\label{Qperiodicity}
\ee  

The transfer matrix eigenvalues for the case considered here are therefore given by 
\be
\Lambda^{(\frac{1}{2},s)}(u) = h(u) {Q(u + (q-1)\eta)\over Q(u)} 
+ h(-u+(q-1) \eta) {Q(u -(q-1)\eta)\over Q(u)}  \,.
\label{tq} 
\ee
We recall that in \cite{Murganspins}, where $\eta = \frac{i \pi}{q}$ for the same choice of boundary parameters, 
the form (\ref{tq}) was found to hold as is the case here when $\eta (=i\pi\frac{r}{q})$. The $h(u)$ function for any odd integer values of $r$ given by (\ref{hoddr}), 
is the same as the one found in \cite{Murganspins}, where only $r = 1$ case was considered. However, from (\ref{hevenr}), we see that the $h(u)$ function for
even integer values of $r$ differ a little from the one given by (\ref{hoddr}), with the former lacking the spin dependent term ($2sN$),  while the latter is not. This is due to the fact that the $f(u)$ function (see (\ref{f0})-(\ref{f1reven})) 
for these cases are indeed different as well. 

The function $Q(u)$ however has the same structure as in \cite{Murganspins} and is given by,
\be
Q(u) = \prod_{j=1}^{M} 
\sh \left( {1\over 2}(u - u_{j}) \right)
\sh \left( {1\over 2}(u + u_{j} - (q-1)\eta) \right) \,,
\label{Q}
\ee 
where
\be
M=2sN+q-1
\,, \label{M}
\ee
It can be noted that $Q(u)$ has the $2i\pi$-periodicity and the crossing symmetry, 
$Q(-u+ (q-1)\eta) = Q(u)$. $\Large\{u_{j}\Large\}$ are the Bethe roots which are also the zeros of $Q(u)$. We remark that $\alpha$ can be any one from $\{\alpha_{\pm}\}$ 
and $\beta$ can be any one from $\{\beta_{\pm}\}$.  
For rescaling purpose, if one divides (\ref{tq}) by the factor $g^{(\frac{1}{2},s)}(u)^{2N} = \left[\prod_{k=1}^{2s-1} \sh(u+(s-k+\frac{1}{2})\eta)\right]^{2N}$, 
(\ref{tq}) assumes the following familiar form (see also \cite{Murganspins}) in terms of the eigenvalues 
$\tilde \Lambda^{(\frac{1}{2},s)}(u) = \frac{\Lambda^{(\frac{1}{2},s)}(u)}{g^{(\frac{1}{2},s)}(u)^{2N}}$,
\be
\tilde \Lambda^{(\frac{1}{2},s)}(u) = 
\tilde h(u) {Q(u+(q-1)\eta)\over Q(u)} +
\tilde h(-u+(q-1) \eta) {Q(u -(q-1)\eta)\over Q(u)} \,,
\label{TQ3}
\ee
where now
\be
\tilde h(u) &=& (-1)^{2sN} 4\sh^{2N}(u+(s+\frac{1}{2})\eta) 
{\sh(2u+2\eta)\over \sh(2u+\eta)}\non\\
&\times&  \ch u\ch(u-\eta) (\sh u+(-1)^{2sN}i\ch\beta)\sh(u-\alpha){\ch\left({1\over 2}(u+\alpha+\eta) \right)\over 
 \ch\left({1\over 2}(u-\alpha-\eta)\right)}
\label{hoddrtilde}
\ee
for odd integer values of $r$ and
\be
\tilde h(u) &=& -4\sh^{2N}(u+(s+\frac{1}{2})\eta) 
{\sh(2u+2\eta)\over \sh(2u+\eta)}\non\\
&\times&  \ch u\ch(u-\eta) (\sh u-i\ch\beta)\sh(u-\alpha){\ch\left({1\over 2}(u+\alpha+\eta) \right)\over 
 \ch\left({1\over 2}(u-\alpha-\eta)\right)}
\label{hevenrtilde}
\ee
for even integer values of $r$.
The analyticity of $\tilde \Lambda^{(\frac{1}{2},s)}(u)$ gives the following 
Bethe ansatz equations,
\be
{\tilde h(u_{j})\over \tilde h(-u_{j}+(q-1)\eta)} = 
-{Q(u_{j}-(q-1)\eta)\over Q(u_{j}+(q-1)\eta)} \,, 
\qquad j = 1 \,, \ldots \,, M \,.
\label{BAeqs}
\ee
The above solution is confirmed numerically for small values of $N$ and $q$ for $s = 1/2\,,1$ and $3/2$. Finally, we remark that
the $h(u)$ given by (\ref{hoddr}) and (\ref{hevenr}) and therefore the $\tilde h(u)$ given by (\ref{hoddrtilde}) and (\ref{hevenrtilde}) are found largely by 
trial and error and are not the only solutions. Other solutions for $h(u)$ exist for which the number of Bethe roots $M$ may also be different. The $h(u)$ functions given here
yield the smallest $M$ value among other functions we tested. We find this beneficial in fascilitating numerical works.   

\subsection{Case 2 : Arbitrary $\alpha_{+}$ and $\alpha_{-}$}

In this section, we shall take $\alpha_{+}$ and $\alpha_{-}$ to be arbitrary while setting $\beta_{\pm} = \eta\,, \theta_{+} = \theta_{-} =$ arbitrary. 
Our choice of boundary parameters for the present case, like for case 1, will make the discriminants of the corresponding quadratic equations (\ref{quadratic}) perfect squares
so that the factorization involving the function $h(u)$ as in (\ref{producth1}) can be readily accomplished.  
As in case 1, we rely on the functional relations (\ref{funcrltn}) satisfied by $\Lambda^{(\frac{1}{2},s)}(u)$ which we recast as $\det {\cal M} = 0$. The matrix
$\cal M$ for this case has the same structure as (\ref{calMalpha}) and yields the properties (\ref{cond0})-(\ref{cond2}) for the yet to be determined $h(u)$. Following the
steps outlined for the previous case, we obtain the following solutions for the function $h(u)$:
\be
h(u) &=& (-1)^{2sN} 4\left[\prod_{k=0}^{2s-1}\sh(u+(s-k+\frac{1}{2})\eta)\right]^{2N} 
{\sh(2u+2\eta)\over \sh(2u+\eta)}\non\\
&\times&  \ch (u+\eta)\ch(u-\eta) \sh (u+(-1)^{2sN}\alpha_{+})\sh(u-\alpha_{-})\non\\
&\times& {\ch\left({1\over 2}(u+\alpha_{-}+\eta) \right)\over 
 \ch\left({1\over 2}(u-\alpha_{-}-\eta)\right)}{\ch\left({1\over 2}(u-(-1)^{2sN}\alpha_{+}+\eta) \right)\over 
 \ch\left({1\over 2}(u+(-1)^{2sN}\alpha_{+}-\eta)\right)}
\label{hoddraa}
\ee
for odd integer values of $r$ and
\be
h(u) &=& -4\left[\prod_{k=0}^{2s-1}\sh(u+(s-k+\frac{1}{2})\eta)\right]^{2N} 
{\sh(2u+2\eta)\over \sh(2u+\eta)}\non\\
&\times&  \ch (u+\eta)\ch(u-\eta) \sh (u-\alpha_{+})\sh(u-\alpha_{-})\non\\
&\times& {\ch\left({1\over 2}(u+\alpha_{-}+\eta) \right)\over 
 \ch\left({1\over 2}(u-\alpha_{-}-\eta)\right)}{\ch\left({1\over 2}(u+\alpha_{+}+\eta) \right)\over 
 \ch\left({1\over 2}(u-\alpha_{+}-\eta)\right)}
\label{hevenraa}
\ee
for even integer values of $r$. 
The structure of matrix $\cal M$ suggests the same form for its null eigenvector as for the previous case. Consequently, the $Q(u)$ function is identical in form 
to (\ref{Q}) but with a different $M$\footnote{The $M$ here is different (smaller) than the corrresponding $M$ given in \cite{Murganspins}. This suggests that
the $h(u)$ presented here in (\ref{hoddraa}) and (\ref{hoddrtildeaa}), despite only being slightly different to the one given in \cite{Murganspins}, is more suitable 
as far as numerical works are concerned, since the resulting Bethe ansatz equations give less number of Bethe roots for each energy level.}, namely
\be
M = 2sN + q + 1\,.
\label{Malpha}
\ee  
The $T-Q$ equation and the Bethe ansatz equations are therefore given by (\ref{tq}) (and by (\ref{TQ3}) after the usual rescaling) and (\ref{BAeqs}) respectively,
where in (\ref{TQ3}) and (\ref{BAeqs}), the function $\tilde h(u)$ is given by,
\be
\tilde h(u) &=& (-1)^{2sN} 4\sh^{2N}(u+(s+\frac{1}{2})\eta) 
{\sh(2u+2\eta)\over \sh(2u+\eta)}\non\\
&\times&  \ch (u+\eta)\ch(u-\eta) \sh (u+(-1)^{2sN}\alpha_{+})\sh(u-\alpha_{-})\non\\
&\times& {\ch\left({1\over 2}(u+\alpha_{-}+\eta) \right)\over 
 \ch\left({1\over 2}(u-\alpha_{-}-\eta)\right)}{\ch\left({1\over 2}(u-(-1)^{2sN}\alpha_{+}+\eta) \right)\over 
 \ch\left({1\over 2}(u+(-1)^{2sN}\alpha_{+}-\eta)\right)}
\label{hoddrtildeaa}
\ee
for odd integer values of $r$ and
\be
\tilde h(u) &=& -4\sh^{2N}(u+(s+\frac{1}{2})\eta) 
{\sh(2u+2\eta)\over \sh(2u+\eta)}\non\\
&\times&  \ch (u+\eta)\ch(u-\eta) \sh (u-\alpha_{+})\sh(u-\alpha_{-})\non\\
&\times& {\ch\left({1\over 2}(u+\alpha_{-}+\eta) \right)\over 
 \ch\left({1\over 2}(u-\alpha_{-}-\eta)\right)}{\ch\left({1\over 2}(u+\alpha_{+}+\eta) \right)\over 
 \ch\left({1\over 2}(u-\alpha_{+}-\eta)\right)}
\label{hevenrtildeaa}
\ee
for even integer values of $r$. As before, we see the presence of the $2sN$ term in $h(u)$ and $\tilde h(u)$ when $r$ assumes odd integer values. 
The solution is also confirmed numerically for small values of $N$ and $q$ for s = 1/2, 1 and 3/2.

\subsection{Case 3 : Arbitrary $\beta_{+}$ and $\beta_{-}$}

As the final case, we consider the following combination of boundary parameters: $\beta_{\pm}$ arbitrary, $\alpha_{\pm} = \eta$, $\theta_{+} = \theta_{-} =$ arbitrary.
The corresponding $\cal M$ matrix that gives the functional relation (\ref{funcrltn}) when its determinant vanishes is,

$\cal M =$
\be
\left(
\begin{array}{cccccccc}
    \Lambda^{(\frac{1}{2},s)}(u) & -h(u) & 0  & \ldots  & 0 & -h(-u- \eta)  \\
    -h(-u-(p+1)\eta) & \Lambda^{(\frac{1}{2},s)}(u+p\eta) & -h(u+p \eta)  & \ldots  & 0 & 0  \\
    \vdots  & \vdots & \vdots & \ddots 
    & \vdots  & \vdots    \\
   -h(u+p^{2} \eta)  & 0 & 0 & \ldots  & -h(-u-(p^{2}+1) \eta) &
    \Lambda^{(\frac{1}{2},s)}(u+p^{2}\eta)
\end{array} \right)  \,,
\label{calMbeta}\non \\
\ee
where $p + 1 = q$. This is accomplished (for odd integer $r$ values) if $h(u)$ satisfies
\be
h(u + 2 i \pi) = h \left(u +2q\eta \right) &=& h(u) \,,
\label{cond0beta} \\
h(u+(q+1)\eta)\ h(-u-\eta) &=& \delta^{(s)}(u) \,, \label{cond1beta} \\
\prod_{j=0}^{q-1} h(u+2j\eta) + \prod_{j=0}^{q-1} h(-u-(2j+1)\eta) &=& f(u) 
\,. \label{cond2beta} 
\ee
The above conditions yield the following as a solution for $h(u)$, 
\be
h(u) &=& (-1)^{2sN} 4\left[\prod_{k=0}^{2s-1}\sh(u+(s-k+\frac{1}{2})\eta)\right]^{2N} 
{\sh(2u+2\eta)\over \sh(2u+\eta)}\non\\
&\times& \sh(u-\eta)\sh(u+\eta)(\ch u-i\sh\beta_{-})(\ch u+(-1)^{2sN}i\sh\beta_{+})
\label{hbetamp}
\ee
The $T-Q$ equation for the transfer matrix eigenvalues now is given by
\be
\Lambda^{(\frac{1}{2},s)}(u) = h(u) {Q(u + (q-1)\eta)\over Q(u)} 
+ h(-u - \eta) {Q(u -(q-1)\eta)\over Q(u)}  \,.
\label{eigenvaluesbeta} 
\ee
Due to the common factor $g^{(\frac{1}{2},s)}(u)^{2N}$, and using the crossing symmetry $g^{(\frac{1}{2},s)}(u)=\pm g^{(\frac{1}{2},s)}(-u-\eta)$, the
rescaled eigenvalues of $\tilde t^{(\frac{1}{2},s)}(u)$ are given by	
\be
\tilde \Lambda^{(\frac{1}{2},s)}(u) = 
\tilde h(u) {Q(u+(q-1)\eta)\over Q(u)} +
\tilde h(-u-\eta) {Q(u -(q-1)\eta)\over Q(u)} \,,
\label{TQ4}
\ee
where 
\be
\tilde h(u) &=& (-1)^{2sN} 4\sh^{2N}(u+(s+\frac{1}{2})\eta) 
{\sh(2u+2\eta)\over \sh(2u+\eta)}\non\\
&\times& \sh(u-\eta)\sh(u+\eta)(\ch u-i\sh\beta_{-})(\ch u+(-1)^{2sN}i\sh\beta_{+})
\label{hbetamptilde}
\ee
and as in the previous cases, the form of the null eigenvector of the matrix $\cal M$ gives the following for the $Q(u)$ function,
\be
Q(u) = \prod_{j=1}^{M} 
\sh \left( {1\over 2}(u - u_{j}) \right)
\sh \left( {1\over 2}(u + u_{j} + \eta) \right) \,,
\label{Qbeta}
\ee 
which satisfies $Q(u + 2i\pi) = Q(u)$ and $Q(-u-\eta) = Q(u)$, and
\be
M=2sN+q-1
\,. \label{Mbeta}
\ee   
Finally, the analyticity of $\tilde \Lambda^{(\frac{1}{2},s)}(u)$ yields the Bethe ansatz equations,
\be
{\tilde h(u_{j})\over \tilde h(-u_{j}-\eta)} = 
-{Q(u_{j}-(q-1)\eta)\over Q(u_{j}+(q-1)\eta)} \,, 
\qquad j = 1 \,, \ldots \,, M \,.
\label{BAeqsbeta}
\ee 
We stress that the results (\ref{hbetamp})-(\ref{BAeqsbeta}) work only for odd integer $r$ values. 

For even integer $r$ values, the condition $\det \cal M =$ 0 with $\cal M$ given in (\ref{calMbeta}), does not seem to produce
the functional relation ($\ref{funcrltn}$). The difficulty here is to find the function $h(u)$ that satisfies the properties (\ref{cond1beta}) and (\ref{cond2beta}).
Since our method of finding $h(u)$ has been largely by trial and error, it is not clear whether an analogous $h(u)$ can be obtained for this case using the matrix (\ref{calMbeta}). 
Other available choices for $\cal M$ did not help either.        

\section{Hamiltonian, energy eigenvalues and numerical results for open spin-$\frac{1}{2}$ and open spin-$1$ XXZ quantum spin chains.}

Here, we provide numerical support for the completeness of the Bethe ansatz solutions given in the previous section. We stress that these numerical studies are not a 
substitute for a proof for completeness of the solutions, but serve only as numerical verification. More specifically, we compute the energy eigenvalues of both open 
spin-$\frac{1}{2}$ and open spin-$1$ XXZ spin chains. The complete energy levels and the Bethe roots used in the computations are tabulated in Tables 1 and 2.

\subsection{s = 1/2 case}

The Hamiltonian for the open spin-$1/2$ XXZ quantum spin chain is given by \cite{dVGR, GZ}

\be
{\cal H} &=& {1\over 2}\sum_{n=1}^{N-1}\left( 
\sigma_{n}^{x}\sigma_{n+1}^{x}+\sigma_{n}^{y}\sigma_{n+1}^{y} 
+\ch \eta\ \sigma_{n}^{z}\sigma_{n+1}^{z}\right) \non \\
&+& {1\over 2}\sh \eta \Big[ 
\cth \alpha_{-} \tnh \beta_{-}\sigma_{1}^{z}
+ \csch \alpha_{-} \sech \beta_{-}\big( 
\ch \theta_{-}\sigma_{1}^{x} 
+ i\sh \theta_{-}\sigma_{1}^{y} \big) \non \\
& & \quad -\cth \alpha_{+} \tnh \beta_{+} \sigma_{N}^{z}
+ \csch \alpha_{+} \sech \beta_{+}\big( 
\ch \theta_{+}\sigma_{N}^{x}
+ i\sh \theta_{+}\sigma_{N}^{y} \big)
\Big]  \,, \label{Hamiltonianshalf} 
\ee
where $\sigma^{x} \,, \sigma^{y} \,, \sigma^{z}$ are the
standard Pauli matrices, $\eta$ is the bulk anisotropy parameter,
$\alpha_{\pm} \,, \beta_{\pm} \,, \theta_{\pm}$ are arbitrary boundary
parameters, and $N$ is the number of spins.

For illustration purpose, we compute the energy eigenvalues of (\ref{Hamiltonianshalf}) for a particular case where the two arbitrary boundary parameters are 
$\alpha_{-}$ and $\alpha_{+}$, making use of the results found in Sec 3.2. The steps here can be repeated for any other desired combinations of boundary parameters. The Hamiltonian (\ref{Hamiltonianshalf}) 
is related to the first derivative of the transfer matrix, $\tilde t^{(\frac{1}{2},\frac{1}{2})}(u)$\cite{Sklyanin},
\be
{\cal H} = c^{(\frac {1}{2})}_{1} {d \over du} \tilde t^{(\frac{1}{2},\frac{1}{2})}(u) \Big\vert_{u=0} 
+ c^{(\frac {1}{2})}_{2} \id \,,
\label{firstderivative}
\ee
where
\be
c^{(\frac {1}{2})}_{1} &=& -{1\over 16 \sh \alpha_{-} \ch \beta_{-}
\sh \alpha_{+} \ch \beta_{+} \sh^{2N-1} \eta 
\ch \eta} \,, \non \\
c^{(\frac {1}{2})}_{2} &=& - {\sh^{2}\eta  + N \ch^{2}\eta\over 2 \ch \eta} 
\,,
\label{cees}
\ee 
and $\id$ is the identity matrix. For $s=1/2$, $t^{(\frac{1}{2},\frac{1}{2})}(u)=\tilde t^{(\frac{1}{2},\frac{1}{2})}(u)$. Moreover, (\ref{firstderivative}) 
implies that the energy eigenvalues are given by
\be
E = c^{(\frac {1}{2})}_{1} {d \over du} \tilde \Lambda^{(\frac{1}{2},\frac{1}{2})}(u) \Big\vert_{u=0} 
+ c^{(\frac {1}{2})}_{2} \,,
\label{firstderivative2}
\ee
Hence, using the results (\ref{Q})-(\ref{TQ3}) and (\ref{hoddrtildeaa})\footnote{The expression (\ref{energyspinhalf}) is derived for odd integer values of $r$. 
Similar expression for even integer values of $r$ can also be derived.}, 
one arrives at the following result for the energy eigenvalues in terms of Bethe roots $\{u_{j}\}$,
\be
E &=& {1\over 2} \sh\eta\ch{\eta\over 2} \sum_{j=1}^{M}{1\over 
\sh ({1\over 2} u_{j} )\ch ({1\over 2} (u_{j} + \eta) )}
 + {1\over 2}N \ch \eta -{1\over 2}\ch \eta \non \\
&+& {1\over 2}\sh \eta(-\coth \alpha_{-} + (-1)^N\coth \alpha_{+} - (-1)^N \tanh ({\alpha_{+}-(-1)^N\eta\over 2}) \non\\
&+& \tanh({\alpha_{-}+\eta\over 2}))\,.
\label{energyspinhalf}
\ee
where $M = N + q + 1$. The energy eigenvalues computed from the Bethe roots using (\ref{energyspinhalf}) for $N = 4$, which are tabulated in Table 1, 
coincide with those obtained from direct diagonalization of (\ref{Hamiltonianshalf}).

\subsection{s = 1 case}  

As for the spin-$1/2$ case, a brief review of the open spin-$1$ XXZ quantum spin chain is desirable at this point. The Hamiltonian is given by
\be
{\cal H} = \sum_{n=1}^{N-1}H_{n,n+1} + H_{b} \,,
\label{Hamiltonianspin1}
\ee
where the bulk term $H_{n,n+1}$ \cite{FZamod} and the boundary term $H_{b}$ \cite{FNR, IOZ} are given by,
\be 
H_{n,n+1} &=&  \sigma_{n} - (\sigma_{n})^{2}
+ 2 \sh^2 \eta \left[ \sigma_{n}^{z} + (S^z_n)^2
+ (S^z_{n+1})^2 - (\sigma_{n}^{z})^2 \right] \non \\
&-& 4 \sh^2 (\frac{\eta}{2})  \left( \sigma_{n}^{\bot} \sigma_{n}^{z}
+ \sigma_{n}^{z} \sigma_{n}^{\bot} \right) \,, \label{bulkhamiltonianspin1}
\ee 
and
\be 
H_{b} &=& a_{1} (S^{z}_{1})^{2}  + a_{2} S^{z}_{1} 
+  a_{3} (S^{+}_{1})^{2}  +  a_{4} (S^{-}_{1})^{2}  +
a_{5} S^{+}_{1}\, S^{z}_{1}  + a_{6}  S^{z}_{1}\, S^{-}_{1} \non \\
&+& a_{7}  S^{z}_{1}\, S^{+}_{1} + a_{8} S^{-}_{1}\, S^{z}_{1} 
+ (a_{j} \leftrightarrow b_{j} \mbox{ and } 1 \leftrightarrow N) \,,
\label{bound}
\ee
respectively. In (\ref{bulkhamiltonianspin1}) and (\ref{bound}), the following definitions are used,
\be
\sigma_{n} = \vec S_n \cdot \vec S_{n+1} \,, \quad
\sigma_{n}^{\bot} = S^x_n S^x_{n+1} + S^y_n S^y_{n+1}  \,, \quad
\sigma_{n}^{z} = S^z_n S^z_{n+1} \,, 
\ee
where $\vec S$ are the $su(2)$ spin-1 generators and $S^{\pm} = S^{x} \pm i S^{y}$. The coefficients $\{ a_{i} \}$ 
in the boundary terms at site 1 are functions  
of the boundary parameters ($\alpha_{-}, \beta_{-},
\theta_{-}$) and the bulk anisotropy parameter $\eta$. They are given by,
\be
a_{1} &=& \frac{1}{4} a_{0} \left(\ch 2\alpha_{-} - \ch 
2\beta_{-}+\ch \eta \right) \sh 2\eta 
\sh \eta \,,\non \\
a_{2} &=& \frac{1}{4} a_{0} \sh 2\alpha_{-} \sh 2\beta_{-} \sh 2\eta \,, \non \\
a_{3} &=& -\frac{1}{8} a_{0} e^{2\theta_{-}} \sh 2\eta 
\sh \eta \,, \non \\
a_{4} &=& -\frac{1}{8} a_{0} e^{-2\theta_{-}} \sh 2\eta 
\sh \eta \,, \non \\
a_{5} &=&  a_{0} e^{\theta_{-}} \left(
\ch \beta_{-}\sh \alpha_{-} \ch {\eta\over 2} +
\ch \alpha_{-}\sh \beta_{-} \sh {\eta\over 2} \right)
\sh \eta \ch^{\frac{3}{2}}\eta \,, \non \\
a_{6} &=&  a_{0} e^{-\theta_{-}} \left(
\ch \beta_{-}\sh \alpha_{-} \ch {\eta\over 2} +
\ch \alpha_{-}\sh \beta_{-} \sh {\eta\over 2} \right)
\sh \eta \ch^{\frac{3}{2}}\eta \,, \non \\
a_{7} &=&  -a_{0} e^{\theta_{-}} \left(
\ch \beta_{-}\sh \alpha_{-} \ch {\eta\over 2} -
\ch \alpha_{-}\sh \beta_{-} \sh {\eta\over 2} \right)
\sh \eta \ch^{\frac{3}{2}}\eta \,, \non \\
a_{8} &=&  -a_{0} e^{-\theta_{-}} \left(
\ch \beta_{-}\sh \alpha_{-} \ch {\eta\over 2} -
\ch \alpha_{-}\sh \beta_{-} \sh {\eta\over 2} \right)
\sh \eta \ch^{\frac{3}{2}}\eta \,,
\ee
where 
\be
a_{0}= \left[
\sh(\alpha_{-}-{\eta\over 2})\sh(\alpha_{-}+{\eta\over 2})
\ch(\beta_{-}-{\eta\over 2})\ch(\beta_{-}+{\eta\over 2})\right]^{-1} 
\,.
\ee
Similiarly, the coefficients $\{ b_{i} \}$ at 
site $N$ are functions of the boundary parameters ($\alpha_{+}, \beta_{+}, \theta_{+}$) and $\eta$, are given by the following correspondence,
\be
b_{i} = a_{i}\Big\vert_{\alpha_{-}\rightarrow \alpha_{+}, 
\beta_{-}\rightarrow -\beta_{+}, \theta_{-}\rightarrow \theta_{+}} \,.
\ee

Next, we proceed to find an expression for the eigenvalues of the Hamiltonian (\ref{Hamiltonianspin1}) for the case considered in Sec 3.1, namely with two arbitrary 
boundary parameters, one from $\{\alpha_{\pm}\}$ and the other from $\{\beta{\pm}\}$, e.g. $\{\alpha_{+},\beta_{-}\}$, etc. We set $\theta_{-} = \theta_{+}= \theta$, 
where $\theta$ is arbitrary. The anisotropy parameter $\eta$ is set to be $\eta = i\pi\frac{r}{q}$.
The energy eigenvalues in terms of the rescaled transfer matrix eigenvalues $\tilde \Lambda^{(1,1)}(u)$ is given by, 
\be
E = c^{(1)}_{1} {d \over du} \tilde \Lambda^{(1,1)}(u) \Big\vert_{u=0} 
+ c^{(1)}_{2}\,,
\label{energyfirstderivatives1}
\ee
where
\be 
c^{(1)}_{1}&=&\ch \eta \Big\{ 16 [\sh 2\eta \sh \eta]^{2N} \sh 3\eta 
\sh(\alpha_{-}-{\eta\over 2})\sh(\alpha_{-}+{\eta\over 2})
\ch(\beta_{-}-{\eta\over 2})\ch(\beta_{-}+{\eta\over 2})\non \\
&\times& \sh(\alpha_{+}-{\eta\over 2})\sh(\alpha_{+}+{\eta\over 2})
\ch(\beta_{+}-{\eta\over 2})\ch(\beta_{+}+{\eta\over 2})\Big\}^{-1}
\label{c1sone}
\,,
\ee
\be 
c^{(1)}_{2}&=& -{a_{0}\over 4}b\ch\eta - (N-1)(4+\ch 2\eta) + 2 N \ch^{2}\eta \non \\
&-& {\sh\eta\over 2d}\Big\{-2\ch 2\alpha_{+}\Big(\ch\eta (3+7\ch 2\eta +\ch 4\eta)+\ch 2\beta_{+}(4+5\ch 2\eta+2\ch 4\eta)\Big)\non \\
&+& 2\ch \eta\Big(\ch 2\beta_{+}(3+7\ch 2\eta +\ch 4\eta)+\ch\eta (5+3\ch 2\eta +3\ch 4\eta)\Big)\Big\}\non \\
&-& {\sh 2\eta\over 2d}\Big\{\ch 2\beta_{+}(2+4\ch \eta \ch 3\eta)+\ch \eta (5\ch 2\eta +\ch 4\eta)-2\ch 2\alpha_{+}\Big(1+\ch 2\eta \non \\
&+& \ch 2\beta_{+}(\ch \eta +2\ch 3\eta)+\ch 4\eta\Big)\Big\}
\label{c2spin1}\,.
\ee
\noindent In (\ref{c2spin1}), $b$ and $d$ are given by
\be
b = 2\big(-\ch 2\beta_{-}-\ch^{3}\eta + \ch 2\alpha_{-}(1+\ch 2\beta_{-}\ch\eta)\big)  
\ee
and
\be
d = -4\sh 3\eta \sh(\alpha_{+}+{\eta\over 2})\sh(\alpha_{+}-{\eta\over 2})
\ch(\beta_{+}+{\eta\over 2})\ch(\beta_{+}-{\eta\over 2})
\ee
The rescaled spin-$1$ transfer matrix eigenvalues are given by\footnote{Following \cite{FNR}, the rescaled factor $\gamma$
is introduced.}
\be
\tilde \Lambda^{(1,1)}(u) &=& \gamma \Lambda^{(1,1)}(u)
\label{fh2}
\ee
where
$\gamma = {\sh(2u) \sh(2u+2\eta)\over [\sh u 
\sh(u+\eta)]^{2N}}$ and $\Lambda^{(1,1)}(u)$ is given by the result from fusion hierarchy (\ref{hierarchy}). 
The analytic form of the energy eigenvalues in terms of Bethe roots $\Large\{u_{k}\Large\}$ then follows from (\ref{energyfirstderivatives1}), 
\be
E &=& \frac{1}{2}\sh(2\eta)\sh(\eta)\sum_{k=1}^{M}\frac{1}{\sh (\frac{1}{2}(u_{k}+{3\eta\over 2}))\sh (\frac{1}{2}(u_{k} - {\eta\over 2}))} 
+ c^{(1)}_{1}(A'(0)+B'(0)-C'(0))\non\\
&+& c^{(1)}_{2}\,,
\label{energyspinone1}
\ee
where
\be
A(u) &=& \tilde{\tilde h}(u+{\eta\over 2})\tilde{\tilde h}(u-{\eta\over 2})\non \\
B(u) &=& \tilde{\tilde h}(-u+(q-\frac{1}{2})\eta)\tilde{\tilde h}(u+{\eta\over 2})\non \\
C(u) &=& -\gamma\delta^{(1)}(u - \frac{\eta}{2})\non \\
\tilde{\tilde h}(u) &=& \sh(2u+\eta)\tilde{h}(u) \,.
\ee
We recall that $M = 2N + q - 1$. The expression (\ref{energyspinone1}) is derived here for even positive integer values of $r$. Similar result can be derived when
$r$ assumes odd positive integers. This can then be used in the 
computation of complete energy levels from the Bethe roots given by (\ref{BAeqs}). 
We tabulate the energies computed from the Bethe roots $\Large\{u_{k}\Large\}$, using (\ref{energyspinone1}) for some selected values of $N\,, q\,, r$ (therefore $\eta$) and the boundary 
parameters $\Large\{\alpha_{\pm}\,, \beta_{\pm}\,, \theta_{\pm}\Large\}$ in Table 2. These numerical 
results provide support to and illustrate the completeness of the Bethe ansatz equations, (\ref{BAeqs}). We have verified that the energies given in Table 2 
coincide with those obtained from direct diagonalization of the open spin-$1$ XXZ chain Hamiltonian (\ref{Hamiltonianspin1}).

\section{Conclusion}\label{sec:conclude}

Bethe ansatz solutions of an open spin-$s$ XXZ quantum spin chain with nondiagonal boundary terms, derived from certain functional relations which the ``fundamental'' transfer matrices, 
$t^{(\frac{1}{2},s)}(u)$ obey at roots of unity are revisited. The solutions given here include a wider class of anisotropic parameter $\eta$, namely 
$\eta = i \pi\frac{r}{q}$, where $r$ and $q$ are positive integers with $q$ assuming the odd integer values and $\frac{r}{q}$ corresponds to irreducible fractions. 
As far as we know, Bethe ansatz solution for such a case for open spin-$s$ XXZ chain has not been reported, except for the $s=\frac{1}{2}$ case in \cite{CYSW}. 
These considerations have motivated the present work. The solutions given here are for cases with any two arbitrary boundary parameters from the $\{\alpha_{\pm}\,,\beta_{\pm}\}$ set, while
the remaining ones are fixed to some values. These solutions have been checked numerically for chains of length up to $N = 5$.
Numerical support for the completeness of the Bethe ansatz solutions (using $s = 1/2$ and $s = 1$ as examples) are provided in Tables 1 and 2, where all $(2s+1)^N$ eigenvalues  
are given.

There remain problems that are worth investigating. It would be interesting to similarly investigate the solutions of the open quantum 
spin chain with alternating spins with nondiagonal boundary terms. One could also attempt to study the thermodynamics of such a model. Furthermore, Bethe ansatz solutions
for open spin-$1$ XXZ quantum spin chain can be used to investigate the supersymmetric sine-Gordon model with boundaries via their nonlinear integral equations. 
It would be desirable to use the solution presented here to carry out such an analysis. We hope to address these questions in future.

\section*{Acknowledgements}

R.M. would like to thank Saginaw Valley State University for the Faculty Research Grant Award. The authors also deeply thank the referee for very 
constructive suggestions and comments that helped to improve the presentation of the paper.

\newpage
\begin{table}[htb] 
	    \centering
	    \begin{tabular}{|c|c|c|}\hline
	     $E$ &  Bethe roots, $\Large\{u_{k}\Large\}$\\
	      \hline
-4.56711 & 0.475167 + 0.000593 i, 0.475167 - 1.25723 i, 0.057772 + 
 1.88496 i,\\& 0.057772 + i$\pi$, - 2.19745 i, - 
 1.70664 i, - 2.68126 i,\\& 0.314088 i, - 0.87 i, 
 1.57187 i\\	      
-4.34568 & 0.405517 + 0.666815 i, 0.405517 - 1.92345 i, 0.403252 - 
 0.628319 i,\\& 0.0569468 - 2.70038 i, 0.0569468 + 1.44374 i, 
 2.36338 i,\\& - i$\frac{\pi}{2}$, - 1.70664 i, - 2.82866 i, - 
 0.386637 i\\
-3.05199 & 0.693961 - 2.18827 i, 0.693961 + 0.931636 i, - 0.824282 i, 
 0.45 i,\\& - 0.386637 i, - 1.96144 i, - 1.57051 i, 
 1.54118 i,\\& - 2.8309 i, 2.80606 i\\
-2.38474 & 0.717734 + 0.933002 i, 0.717734 - 2.18964 i, - 0.323985 i, - 
 2.01785 i,\\& - 1.56819 i, - 0.87 i, - 1.70664 i, 
 1.57883 i,\\& 2.82555 i, - 2.70265 i\\
-2.17816  & 0.722701 - 2.18991 i, 0.722701 + 0.933271 i, 0.317914 i, - 
 0.949003 i,\\& - 1.70664 i, - 2.81586 i, 2.19787 i,  
 0.767594 i,\\& 1.44577 i, - 0.386637 i\\     
-0.994085 & 0.590036 + 2.51327 i, 0.572252 - 0.628319 i, - 0.386637 i,\\& 
 - 0.852939 i, 0.666397 i, - 1.70664 i, 0.312972 i,\\& 
 1.57986 i, 2.81222 i, 1.5305 i\\
                       -0.603975 & 0.602144 + 2.51327 i, 0.585957 - 0.628319 i, - 1.96477 i,\\& - 
 0.87 i, 0.45 i, - 0.335719 i, - 1.56682 i,\\& 
 2.82363 i, 1.58342 i, 1.46666 i\\                       	
                      -0.243163 & 0.609459 + 2.51327 i, 0.594107 - 0.628319 i, 0.322076 i, - 
 1.70664 i,\\& - 0.952266 i, - 0.386637 i, 0.723371 i, - 
 2.80224 i,\\& 2.19738 i, 1.46531 i\\
\hline
		  \end{tabular}
		  \caption[xxx]{\parbox[t]{0.8\textwidth}{
		  The complete set of $2^N$ energy levels and corresponding Bethe roots 
		  for  
		  $N=4\,, s=1/2\,, \eta = i 7\pi/5\,,
		  \alpha_{-}=0.45 i\,, \beta_{-}=\eta\,, 
		  \theta_{-}=0.54\,, \alpha_{+}=0.87 i\,, 
		  \beta_{+}=\eta\,, \theta_{+}=0.54 
		  $}
		  }
		 \label{table:energiesM}
\end{table}  
\begin{table}[htb] 
	    \centering
	    \begin{tabular}{|c|c|}\hline
	     $E$ (continued) &  Bethe roots ${u}_{j}$ (continued)\\
	      \hline
     
1.14152 - 0.195122 i & 0.35837 - 2.71807 i, 0.330039 + 2.30814 i, 0.276567 + 
 0.656084 i,\\& 0.087861 - 0.795393 i, 0.027171 - 
 0.308006 i, 0.015303 - 1.55285 i,\\& 0.001193 + 
 2.82864 i, 0.000753 - 2.847 i, - 0.386637 i, 0.45 i\\  
1.14152 + 0.195122 i & 0.35837 + 1.46144 i, 0.330039 + 2.71841 i, 0.276567 - 
 1.91272 i,\\& 0.087861 - 0.461244 i, 0.027171 - 
 0.948631 i, 0.015303 + 0.296211 i,\\& 0.001193 + 
 2.19791 i, 0.000753 + 1.59036 i, 0.45 i, - 0.386637 i\\
1.6454 - 0.036207 i & 0.37612 - 2.71959 i, 0.347861 + 2.30928 i, 0.266598 + 
 0.632143 i,\\& 0.126391 - 0.803703 i, 0.021899 + 
 1.52824 i, 0.013806 + 0.376725 i,\\& 0.008631 - 
 0.942847 i, 0.000513 + 2.19931 i, 0.45 i, - 0.386637 i\\
1.6454 + 0.036207 i & 0.37612 + 1.46295 i, 0.347861 + 2.71727 i, 0.266598 - 
 1.88878 i,\\& 0.126391 - 0.452934 i, 0.021899 - 2.78488 i, 0.013806 - 
 1.63336 i,\\& 0.008631 - 0.31379 i, 0.000513 + 2.82724 i, - 
 1.70664 i, - 0.87 i\\
1.87399 - 0.362703 i & 0.382144 - 2.74196 i, 0.357472 + 2.28825 i, 0.228038 + 0.649592 i,\\& 
 0.144987 - 0.887371 i, 0.120703 + 0.3564 i, 0.035021 - 2.7448 i,\\& 
 0.000215 - 0.93752 i, 0.000022 + 2.19938 i, 0.45 i, 
 - 0.386637 i\\
1.87399 + 0.362703 i & 0.382144 + 1.48532 i, 0.357472 + 2.7383 i, 0.228038 - 1.90623 i,\\& 
 0.144987 - 0.369266 i, 0.120703 - 1.61304 i, 0.035021 + 1.48816 i,\\& 
 0.000215 - 0.319117 i, 0.000022 + 2.82716 i, - 1.70664 i, 
 - 0.386637 i\\
3.41127 & 0.426274 + 0.768867 i, 0.426274 - 2.0255 i, 0.380283 - 2.48686 i,\\& 
 0.380283 + 1.23022 i, 0.264586 + 2.51327 i, - 0.46653 i,\\& 
 - 0.87 i, - 1.70664 i, 2.19908 i, - 0.942942 i\\
5.63582 & 0.610828 + 2.51327 i, 0.585745 - 0.628319 i, 0.264927 - 2.30695 i,\\& 
 0.264927 + 1.05031 i, 0.239528 + 2.51327 i, - 0.386637 i,\\& 
 - 0.786041 i, 0.45 i, - 0.313583 i, 2.19908 i\\                  
                                                                    
\hline
		  \end{tabular}
                  \end {table}  
\begin{table}[htb] 
	    \centering
	    \begin{tabular}{|c|c|c|}\hline
	     $E$ &  Bethe roots, $\Large\{u_{k}\Large\}$\\
	      \hline
-5.983890 & 0.705185 + 1.409455 i, 0.705185 + 
 3.078533 i, 0.548923 - 1.646975 i,\\& 0.548923 - 0.148219 i, 0.210780 + 
 3.018164 i, 0.210780 + 1.469825 i,\\& -0.367144 i, 2.080271 i, -2.080446 i,\\& -2.407592 i\\	      
-4.833822 - 0.089904 i & 0.565328 + 1.464054 i, 0.560227 + 
 3.079482 i, 0.383486 - 1.400808 i,\\& 0.370909 + 0.713516 i, 0.359969 - 
 2.475472 i, 0.253186 - 0.363528 i\\& 0.103171 + 1.549115 i, 0.000260 + 
 2.081055 i, 0.000123 + 0.285436 i,\\& -2.407592 i\\
-4.833822 + 0.089904 i & 0.565328 + 3.023935 i, 0.560227 + 
 1.408507 i, 0.383486 - 0.394387 i,\\& 0.370909 - 2.508711 i, 0.359969 + 
 0.680276 i, 0.253186 - 1.431667 i,\\& 0.103171 + 
 2.938874 i, 0.000260 + 
 2.406933 i, 0.000123 - 2.080631 i,\\& 0.612397 i\\
-2.835193 - 0.109209 i & 0.577091 + 1.584580 i, 0.454273 - 
 2.807273 i, 0.444406 + 0.861744 i,\\& 0.443204 - 0.977039 i, 0.343279 + 
 2.736768 i, 0.251255 - 1.788773 i,\\& 0.016001 + 0.153832 i, 0.011279 + 
 1.949951 i, 0.001045 + 0.732900 i,\\& -2.407592 i\\
-2.835193 + 0.109209 i & 0.577091 + 2.903409 i, 0.454273 + 
 1.012077 i, 0.444406 - 2.656940 i,\\ & 0.443204 - 0.818156 i, 0.343279 + 
 1.751220 i, 0.251255 - 0.006423 i,\\ & 
 0.016001 - 1.949027 i, 0.011279 + 
 2.538038 i, 0.001045 - 2.528096 i,\\& 0.612397 i\\     
-1.859189 - 0.040090 i & 0.624365 + 3.096684 i, 0.613412 + 
 1.442585 i, 0.469863 - 1.439936 i,\\& 0.313319 - 0.296575 i, 0.171303 + 
 1.588756 i, 0.025225 - 2.380049 i,\\& 0.019867 + 2.114178 i, 0.019825 + 
 0.318497 i, 0.003054 + 0.717989 i,\\& -2.407592 i\\
                       -1.859189 + 0.040090 i & 0.624365 + 1.391305 i, 0.613412 + 
 3.045404 i, 0.469863 - 0.355259 i,\\& 0.313319 - 1.498621 i, 0.171303 + 
 2.899233 i, 0.025225 + 0.584854 i,\\& 0.019867 + 2.373811 i, 0.019825 - 
 2.113692 i, 0.003054 - 2.513185 i,\\& -2.407592 i\\                       	
                      -0.818531 - 0.180442 i & 0.607702 + 1.556139 i, 0.466814 - 
 3.064030 i, 0.444104 + 0.768879 i,\\& 0.415149 - 1.265817 i, 0.343162 - 
 2.450324 i, 0.109869 + 0.650801 i,\\& 0.056399 + 2.447113 i, 0.055560 - 
 2.041603 i, 0.010588 - 2.518368 i,\\& -2.407592 i\\	     
-0.818531 + 0.180442 i & 0.607702 + 2.931849 i, 0.466814 + 
 1.268834 i, 0.444104 - 2.564075 i,\\& 0.415149 - 0.529378 i, 0.343162 + 
 0.655129 i, 0.109869 - 2.445997 i,\\& 0.056399 + 2.040875 i, 0.055561 + 
 0.246407 i, 0.010588 + 0.723172 i,\\& 0.612397 i\\                    
                                                                    
\hline
		  \end{tabular}
		  \caption[xxx]{\parbox[t]{0.8\textwidth}{
		  The complete set of $3^N$ energy levels and corresponding Bethe roots 
		  for  
		  $N=2\,, s=1\,, \eta = i 4\pi/7\,,
		  \alpha_{-}=i \pi/2\,, \beta_{-}=0.651\,, 
		  \theta_{-}=0.386\,, \alpha_{+}=0.734 i\,, 
		  \beta_{+}=\eta\,, \theta_{+}=0.386 
		  $}
		  }
		 \label{table:energiesM}
\end{table}  


\begin{thebibliography}{99}

\bibitem{BR}
V.V. Bazhanov and N.Yu. Reshetikhin, 
``Critical RSOS Models And Conformal Field Theory,''
{\it Int. J. Mod. Phys.} {\bf A4}, 115 (1989).

\bibitem{CLSW}
J. Cao, H.-Q. Lin, K.-J. Shi and Y. Wang, 
``Exact solution of XXZ spin chain with unparallel boundary fields,''  
{\it Nucl. Phys.} {\bf B663}, 487 (2003).

\bibitem{Nep}
R.I. Nepomechie, 
``Solving the open XXZ spin chain with nondiagonal boundary terms at roots of unity,'' 
{\it Nucl. Phys.} {\bf B622}, 615 (2002); 
Addendum, {\it Nucl. Phys.} {\bf B631}, 519 (2002)
[{\tt hep-th/0110116}].

\bibitem{Nep2}
R.I. Nepomechie, 
``Bethe Ansatz solution of the open XXZ chain with nondiagonal boundary terms,'' 
{\it J. Phys.} {\bf A37},  433 (2004) [{\tt hep-th/0304092}].

\bibitem{NepRav}
R.I. Nepomechie and F. Ravanini,
``Completeness of the Bethe Ansatz solution of the open XXZ chain with
nondiagonal boundary terms,'' 
{\it J. Phys.} {\bf A36}, 11391 (2003);
Addendum, {\it J. Phys.} {\bf A37}, 1945 (2004) [{\tt hep-th/0307095}].

\bibitem{YNZ}
W.-L. Yang, R.I. Nepomechie and Y.-Z. Zhang,
``Q-operator and T-Q relation from the fusion hierarchy,'' 
{\it Phys. Lett.} {\bf B633}, 664 (2006)
[{\tt hep-th/0511134}].

\bibitem{YZ1}
W.-L. Yang and Y.-Z. Zhang, 
``On the second reference state and complete eigenstates of the open 
XXZ chain,'' 
{\it JHEP} {\bf 04}, 044 (2007)
[{\tt hep-th/0703222}].

\bibitem{MNS3}
R. Murgan, R.I. Nepomechie and C. Shi,
``Exact solution of the open XXZ chain with general integrable boundary terms at 
roots of unity,'' 
{\it J. Stat. Mech} {\bf P08006} (2006)
[{\tt hep-th/0605223}].

\bibitem{BK}
P. Baseilhac and K. Koizumi,
``A deformed analogue of Onsager's symmetry in the XXZ open spin
chain,''
{\it J. Stat. Mech.} {\bf P10005} (2005)
[{\tt hep-th/0507053}].

\bibitem{BK2}
P. Baseilhac,
``The q-deformed analogue of the Onsager algebra: beyond the Bethe ansatz
approach,''
{\it Nucl. Phys.} {\bf B754}, 309 (2006)
[{\tt math-ph/0604036}].

\bibitem{BK3}
P. Baseilhac and K. Koizumi,
``Exact spectrum of the XXZ open spin chain from the $q$-Onsager 
algebra representation theory,''
{\it J. Stat. Mech.} {\bf P09006} (2007)
[{\tt hep-th/0703106}].

\bibitem{Galleas}
W. Galleas,
``Functional relations from the Yang-Baxter algebra: Eigenvalues of the XXZ model with non-diagonal twisted and open boundary conditions,''
{\it Nucl. Phys.} {\bf B790}, 524 (2008)
[{\tt nlin.SI 0708.0009}].

\bibitem{CRS}
N. Crampe, E. Ragoucy and D. Simon,
``Eigenvectors of open XXZ and ASEP models for a class of non-diagonal boundary conditions,"
{\it J. Stat. Mech.} {\bf P11038} (2010)
[{\tt cond-mat/1009.4119}].

\bibitem{CRS2}
N. Crampe, E. Ragoucy and D. Simon,
``Matrix Coordinate Bethe Ansatz: Applications to XXZ and ASEP models,"
{\it J.Phys.} {\bf A44}, 405003 (2011)
[{\tt cond-mat/1106.4712}].

\bibitem{Niccoli}
G. Niccoli,
``Non-diagonal open spin-1/2 XXZ quantum chains by separation of variables: Complete spectrum and matrix elements of some quasi-local operators,''
{\it J. Stat. Mech.} {\bf P10025} (2012)
[{\tt math-ph 1206.0646}].

\bibitem{FKN}
S. Faldella, N. Kitanine and G. Niccoli,
``Complete spectrum and scalar products for the open spin-1/2 XXZ quantum chains with non-diagonal boundary terms,''
{\it J. Stat. Mech.} {\bf P01011} (2014)
[{\tt math-ph 1307.3960}].

\bibitem{KMG}
N. Kitanine, J.-M. Maillet and G. Niccoli,
``Open spin chains with generic integrable boundaries: Baxter equation and Bethe ansatz completeness from SOV''
{\it J. Stat. Mech.} {\bf P05015} (2014)
[{\tt math-ph 1401.4901}].

\bibitem{CYSW}
J. Cao, W. Yang, K. Shi and Y. Wang,
``Off-diagonal Bethe ansatz solutions of the anisotropic spin-$1/2$ chains with arbitrary boundary fields,''
{\it Nucl. Phys.} {\bf B877}, 152 (2013)
[{\tt cond-mat.stat-mech 1307.2023}].

\bibitem{Doikou}
A. Doikou, 
``Fused integrable lattice models with quantum impurities and open 
boundaries,''
{\it Nucl. Phys.} {\bf B668}, 447 (2003) 
[{\tt hep-th/0303205}].

\bibitem{FNR}
L. Frappat, R.I. Nepomechie and E. Ragoucy,
``Complete Bethe ansatz solution of the open spin-$s$ XXZ chain with general integrable boundary terms,''
{\it J. Stat. Mech.} {\bf P09008} (2007)
[{\tt math-ph/0707.0653}].

\bibitem{BSSG}
T. Inami, S. Odake and Y.-Z. Zhang, 
``Supersymmetric extension of the sine-Gordon theory with integrable 
boundary interactions,''
{\it Phys. Lett.} {\bf B359}, 118 (1995) [{\tt hep-th/9506157}].

\bibitem{BSSG2}
R.I. Nepomechie, 
``The boundary supersymmetric sine-Gordon model revisited,'' 
{\it Phys. Lett.} {\bf B509}, 183 (2001) [{\tt hep-th/0103029}].

\bibitem{BSSG3}
Z. Bajnok, L. Palla and G. Tak\'acs,
``Spectrum of boundary states in N = 1 SUSY sine-Gordon theory,''
{\it Nucl. Phys.} {\bf B644}, 509 (2002) [{\tt hep-th/0207099}].

\bibitem{ANS}
C. Ahn, R.I. Nepomechie and J. Suzuki,
``Finite size effects in the spin-1 XXZ and supersymmetric sine-Gordon models
with Dirichlet boundary conditions,''
{\it Nucl. Phys.} {\bf B767}, 250 (2007)
[{\tt hep-th/0611136}].

\bibitem{MurganSSG}
R. Murgan,
``A note on the IR limit of the NLIEs of boundary supersymmetric sine-Gordon model,"
{\it JHEP} {\bf 09}, 059 (2011)
[{\tt hep-th/1107.1928}].

\bibitem{Murganspins}
R.Murgan,
``Bethe ansatz of the open spin-$s$ XXZ chain with nondiagonal boundary terms,'' 
{\it JHEP} {\bf 04}, 076 (2009)
[{\tt hep-th/0901.3558}].

\bibitem{Sklyanin}
E. K. Sklyanin,
``Boundary conditions for integrable quantum systems,''
{\it J. Phys.} {\bf A21} (1988) 2375;

\bibitem{dVGR}
H.J. de Vega and A. Gonz\'alez-Ruiz, 
``Boundary K-matrices for the
six vertex and the $n(2n-1)$ $A_{n-1}$ vertex models,'' 
{\it J. Phys.} {\bf A26}, L519 (1993) 
[{\tt hep-th/9211114}].

\bibitem{GZ}
S. Ghoshal and A.B. Zamolodchikov, 
``Boundary S-Matrix and Boundary State in Two-Dimensional 
Integrable Quantum Field Theory,''
{\it Int. J. Mod. Phys.} {\bf A9}, 3841 (1994) [{\tt hep-th/9306002}].

\bibitem{fusion}
P.P. Kulish, N.Yu. Reshetikhin and E.K. Sklyanin,
``Yang-Baxter equation and representation theory. I,'' 
{\it Lett. Math. Phys.} {\bf 5} (1981) 393.

\bibitem{fusion2}
L. Mezincescu, R.I. Nepomechie and V. Rittenberg,
``Bethe ansatz solution of the Fateev-Zamolodchikov quantum spin chain with boundary terms,''
{\it Phys. Lett.} {\bf A147}, 70 (1990).

\bibitem{fusion3}
L. Mezincescu and R.I. Nepomechie, ``Fusion procedure for open chains,''
{\it J. Phys.} {\bf A25}, 2533 (1992).

\bibitem{Zhou}
Y.-K. Zhou, 
``Row transfer matrix functional relations for Baxter's eight-vertex 
and six-vertex models with open boundaries via more general 
reflection matrices,''
{\it Nucl. Phys.} {\bf B458}, 504 (1996) [{\tt hep-th/9510095}].

\bibitem{IOZ}
T. Inami, S. Odake and Y.-Z. Zhang, 
``Reflection K matrices of the 19 vertex model and XXZ spin 1 chain 
with general boundary terms,''	
{\it Nucl. Phys.} {\bf B470}, 419 (1996) [{\tt hep-th/9601049}].

\bibitem{FZamod}
A.B. Zamolodchikov and V.A. Fateev, 
``Model factorized S matrix and an integrable Heisenberg chain with spin 1,''
 {\it Sov. J. Nucl. Phys.} {\bf 32}, 298 (1980).

\end{thebibliography}
\end{document}